\documentclass[prl,twocolumn,amsmath,amssymb,superscriptaddress,showpacs]{revtex4-1}
\pdfoutput=1
\usepackage{graphicx}
\usepackage{amsmath}
\usepackage{hyperref}

\hypersetup{colorlinks=true, citecolor=blue, linkcolor=black}

\newcommand{\ud}{\mathrm{d}}

\newcommand{\af }{Joint Laboratory for Attosecond Science, University of Ottawa and National Research Council, 100 Sussex Drive, Ottawa, Canada}
\newcommand{\aff}{Department of Physics, University of New Mexico, Albuquerque, USA}
\newcommand{\afff}{Department of Physics, Texas A \& M University, College Station, USA}
\newcommand{\affff}{Laboratoire d'Optique Appliqu\'e, \'Ecole Nationale Sup\'erieure des Techniques Avanc\'ees -- \'Ecole Polytechnique, Chemin de la Huni\`ere, 91761 Palaiseau Cedex, France}

\begin{document}

\title{Partitioning of the linear photon momentum in multiphoton ionization}

\author{C Smeenk}
\affiliation{\af }
\author{ L Arissian}
\affiliation{\af}
\affiliation{\aff}
\affiliation{\afff}
\author{ B Zhou}
\affiliation{\affff}
\author{A Mysyrowicz}
\affiliation{\affff}
\author{D M Villeneuve}
\affiliation{\af}
\author{A Staudte}
\affiliation{\af}
\author{P B Corkum}
\affiliation{\af}

\begin{abstract}
	The balance of the linear photon momentum in multiphoton ionization is studied experimentally. In the experiment argon and neon atoms are singly ionized by  circularly polarized laser pulses with a wavelength of 800~nm and 1400~nm in the intensity range of $10^{14}  - 10^{15} \ \mathrm{W/cm}^2$. The photoelectrons are measured using velocity map imaging. We find that the photoelectrons carry linear momentum corresponding to the photons absorbed above the field free ionization threshold. Our finding has implications for concurrent models of the generation of terahertz radiation in filaments. 
\end{abstract}


\maketitle

The photon momentum is usually eclipsed by other photon properties, such as energy or angular momentum. At high photon energies the photon momentum is manifest, e.g., in the weak non-dipole contributions to the photo-electron angular distributions \cite{Kraessig1995PRL}. Although the momentum of a single visible photon is exceedingly small, a large number of photons can give rise to macroscopic effects of radiation pressure \cite{Marx1966Nature,Gigan2006Nature}. In multiphoton ionization at relativistic light intensities radiation pressure is known to cause a suppression of recollision \cite{Corkum1993} and therefore the suppression of both non-sequential multiple ionization \cite{Dammasch2001PRA,DiChiara2010PRA}, and high harmonic generation \cite{Walser2000PRL}. Recently, radiation pressure has been invoked as an explanation for the terahertz radiation emitted by laser generated filaments \cite{Cheng2001,Amico2008,Sprangle2004,Zhou2011}. However, the detailed balance of the quantized momentum transfer from the photon field to the electron-ion system in multi-photon ionization (MPI) is both theoretically and experimentally unexplored.

We have studied the photo-electron momentum distribution produced in circularly polarized, infrared, femtosecond laser pulses along the direction of the laser pulse propagation. We observe and quantify the effect of radiation pressure on isolated electrons following multi-photon ionization. We find that the asymptotic electron momentum in the direction of photon propagation corresponds to a fraction of the total momentum absorbed in multi-photon ionization. Specifically, neither the photons driving the quiver motion of the free electron during the laser pulse nor the photons required to lift the electron from the bound state into the continuum contribute to the forward momentum of the free electron after the pulse. We will show that the former is clear from semi-classical physics. The physics underlying the latter has never been considered.

Previous studies of the photon momentum transfer were performed at relativistic intensities and the long pulse regime. In these conditions the ionization potential $I_p$ is negligible and the photo-electron exits the focus in the presence of the high intensity pulse \cite{Moore1995,Meyerhoffer1996}. In our experiment the photo-electron kinetic energy is comparable to $I_p$ and the photoelectron remains in the focus for the duration of the pulse. These conditions are similar to most THz generation experiments \cite{Couairon2007}. 

In multi-photon ionization, the total energy absorbed by the photo-electron is given by the number of photons \cite{Freeman1991} $N \hbar \omega  = I_p + K + U_p$, where $I_p$ is the ionization potential of the atom, $K$ is the drift kinetic energy of the electron measured after the event and $U_p$ is the ponderomotive energy of the electron. In a circularly polarized pulse $U_p = q_e^2 E_0^2/\left(2 m \omega^2\right)$, where $E_0$ and $\omega$ are the light electric field and frequency, respectively. Photoelectron distributions of MPI clearly reveal the quantized nature of the transfer of energy and angular momentum. Likewise, the transferred linear momentum must be conserved $N \hbar k = p_z^e + p_z^i + p^L$ where z is the direction of light propagation and the superscripts denote the momenta of photo-electron, photo-ion, and laser field, respectively.

\begin{figure}
	\includegraphics[width=\columnwidth]{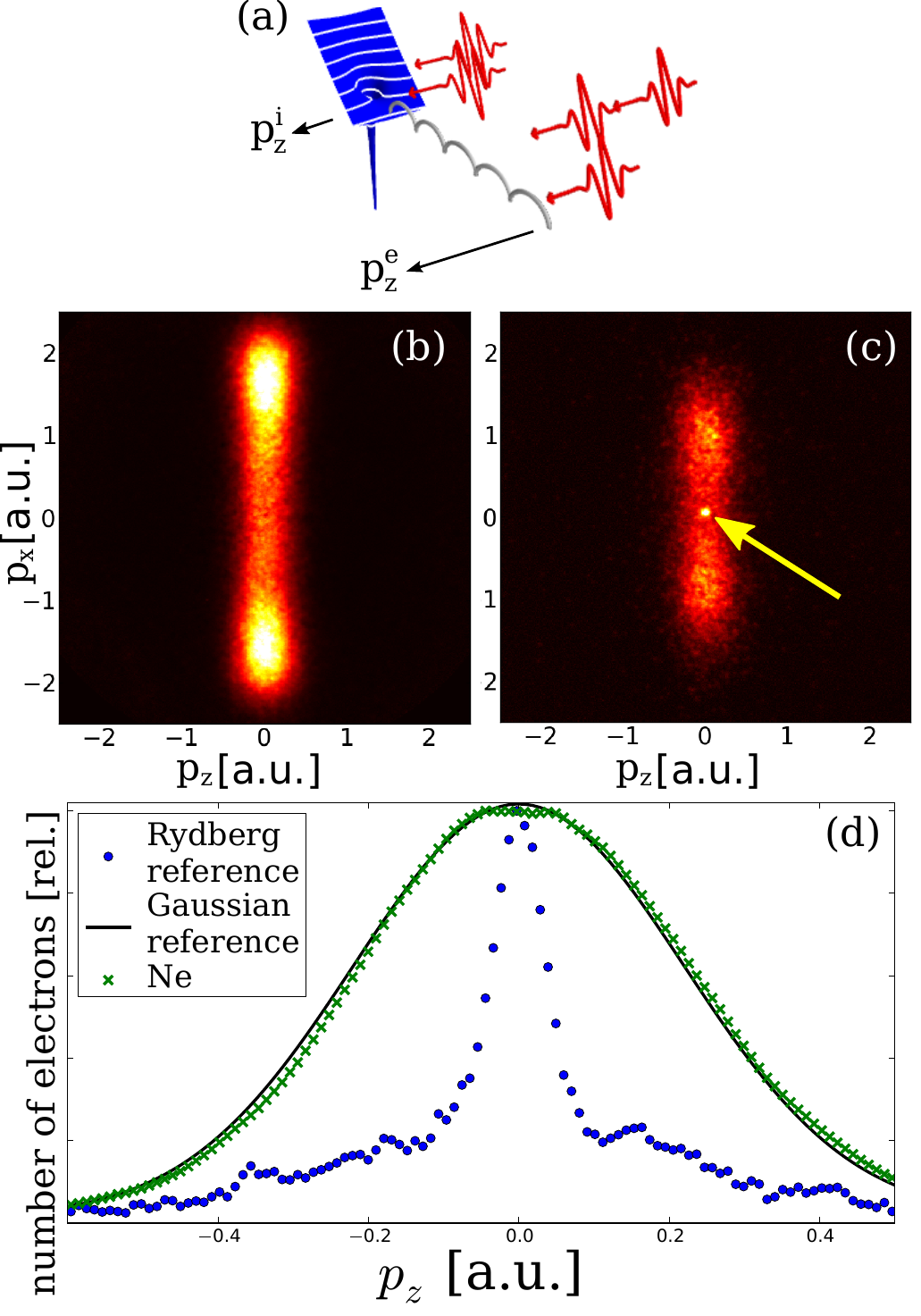}
	\caption{(a) Photoionization of a neutral atom gives a net momentum to the photoelectron $p_z^e = K/c$ and the ion $p_z^i = I_p/c$ in the direction of light propagation. (b) Measured electron momentum spectrum (in atomic units) of Ne in circularly polarized light at 800~nm, $8 \times 10^{14}$ W/cm$^2$. The laser pulse propagates along the z-axis. (c) Background image under the same conditions as in (b) reveals a distinct peak of low energy electrons from laser produced Rydberg atoms that are ionized by the spectrometer DC field (indicated by the arrow). (d) The measured Ne photoelectron distribution (crosses) is compared to the Rydberg reference distribution (dots) and a reference Gaussian distribution centred at $p_z=0$. The centre of the Ne distribution has a net $p_z > 0$.}
	\label{fig:refSpectrum}
\end{figure}

We used circularly polarized laser pulses at 800~nm, 15~fs (intensity FWHM) and 1400~nm, 70~fs to ionize Ar or Ne atoms. The energy of the 800~nm pulse was controlled using a half-wave plate and two germanium plates positioned at Brewster's angle. The energy of the 1400~nm pulse was controlled using a neutral density filter. Our experiment spans the intensity range from $10^{14}-10^{15}$ W cm$^{-2}$. At these intensities and wavelengths, the photo-electron drift velocity is less than 2\% of the velocity of light.

We measure the photo-electron momentum using a velocity map imaging (VMI) spectrometer \cite{Eppink1997}. The spectrometer projects the 3-dimensional electron momentum distribution onto a 2-dimensional imaging detector (Fig.~\ref{fig:refSpectrum}(b)). The detector plane contains the direction of laser propagation (z) and the axis of the gas jet (x). The laser pulses were focused into the super sonic gas jet with a local density of $\approx 10^8$ cm$^{-3}$.  The velocity distribution of the atoms in our pulsed gas jet corresponds to a spread in the electron momentum distribution of $<1.5 \times 10^{-5} \mathrm{a.u.}$ (atomic units - a.u.). 

The nominal spectrometer momentum resolution, determined by its pixel size, is $7 \times 10^{-3}$~a.u. for Ar and $1 \times 10^{-2}$~a.u. for Ne.  However, multi-photon ionization creates a Gaussian wave packet in momentum space along the laser propagation k-direction \cite{Arissian2010} that is much broader than the nominal resolution. A typical wavepacket that we measure is shown as the crosses in Fig.~\ref{fig:refSpectrum}(d). Taking the centroid of this distribution, we can determine a shift between momentum distributions obtained under differing conditions with an accuracy of  $ \sim 2 \times 10^{-3}$~a.u. Since the linear momentum of a single 800~nm photon is $4 \times 10^{-4}$~a.u., we are sensitive to a momentum shift caused by the absorption of $\approx 5$ photons. 

To determine the absolute shift we use electrons extracted from Rydberg states that are populated by multi-photon excitation \cite{Nubbemeyer2008, Eichmann2009} as a reference. The lifetimes of these Rydberg states can easily exceed the duration of our laser pulses. When the laser pulse passes, atoms left in highly excited Rydberg states can be ionized by the spectrometer's DC electric field, resulting in very narrowly distributed, low momentum electrons (see Fig.~\ref{fig:refSpectrum}(c)). The Rydberg states are most efficiently populated in linear light. However, the water vapor in the background gas provides Rydberg electrons even in circular polarized light. We estimate that the electrostatic fields of the spectrometer $0.6 \ \mathrm{kV/cm} \leq E_{stat} \leq 1.3 \ \mathrm{kV/cm}$ can field ionize Rydberg states with $n \geq 20$. 

\begin{figure}
	\includegraphics[width=\columnwidth]{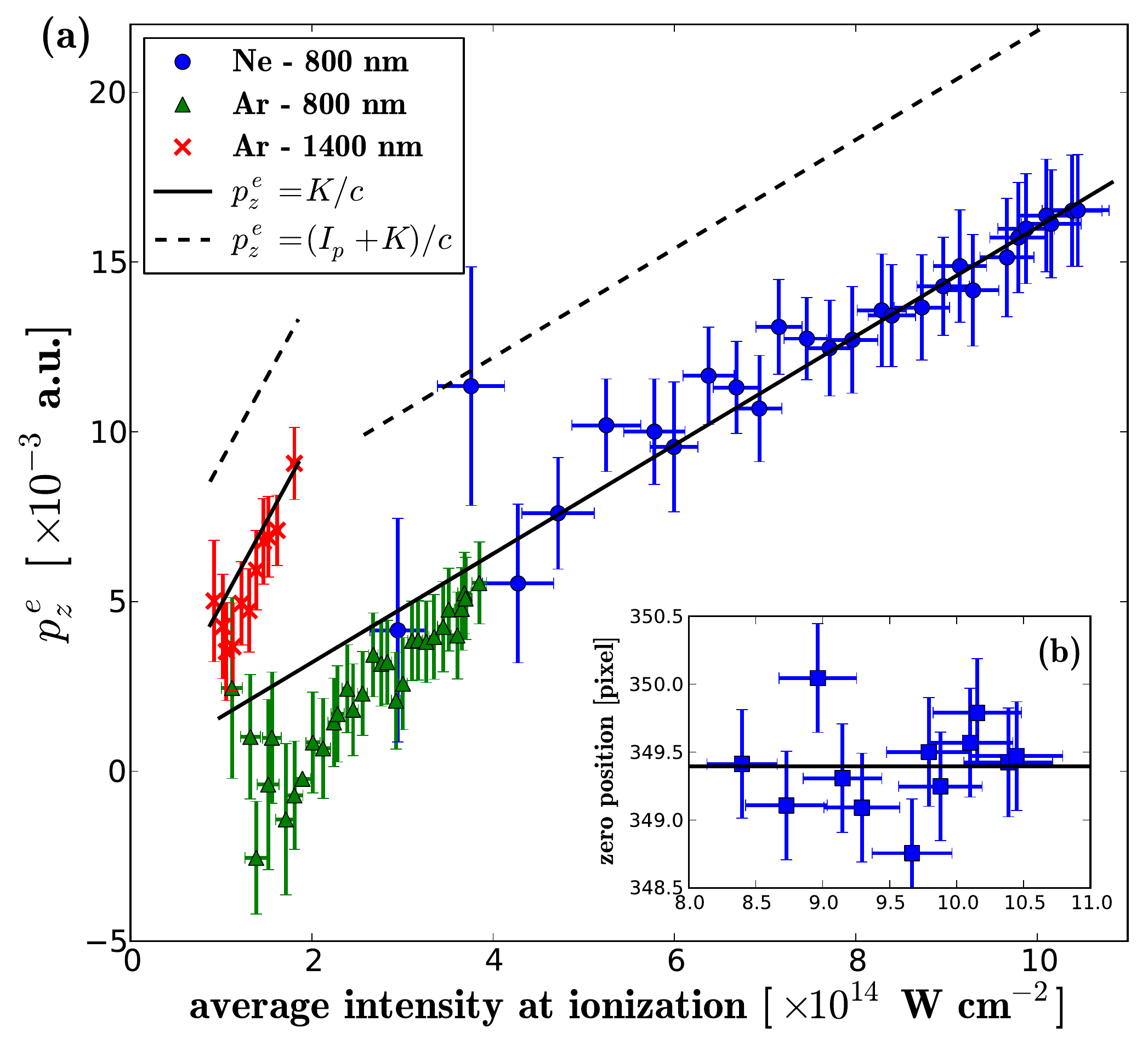}
	\caption{(a) The center of the electron momentum distribution along z as a function of the average intensity at ionization for Ar and Ne at 800~nm and 1400~nm. The intensity at ionization is determined from the electron $p_x$ distribution averaged over the ionization volume (space) and laser pulse length (time; see text for details). The theory lines are given by Eq.\ \eqref{eq:pz}. (b) The zero position of the momentum distribution along the z axis as a function of laser intensity. The position is determined by the center of the Rydberg electron spot indicated by the arrow in Fig.\ \ref{fig:refSpectrum}c. }
	\label{fig:vzDeflection}
\end{figure}

Rydberg electrons provide an unambiguous reference for the zero momentum. The Rydberg atom or molecule has absorbed approximately $N=I_p/(\hbar \omega)$ photons. Since the multi-photon absorption and ionization steps are decoupled in Rydberg atoms, the linear momentum $I_p/c$ is transferred to the center of mass of the atom \cite{Eichmann2009}. Hence, the origin of the Rydberg electron distribution coincides with zero velocity in the lab frame.  By recording many Rydberg electron spectra for a range of laser intensities (shown in Fig.\ \ref{fig:vzDeflection}b) we determine the origin of the momentum distribution with an accuracy of $1 \times 10^{-3}$ a.u.

Our choice of circularly polarized light has two important advantages over linear polarization. It avoids Coulomb focusing \cite{Comtois2005} in the electron momentum distributions and provides a simple measure for the laser intensity. In circular polarization the electron momentum distribution resembles a torus \cite{Smeenk2009} whose radius is proportional to the peak of the laser field experienced by the neutral atoms. Since we are only interested in single ionization this is the appropriate intensity measure \cite{Staudte2009,Akagi2009}.  We determine the light intensity with a precision of 8\% \cite{Smeenk2011}, thereby directly measuring the net number of photons absorbed for a typical electron at that intensity.   

For the data analysis we integrated the measured \mbox{2-D} spectra over x to arrive at a \mbox{1-D} distribution along the axis of laser propagation (the crosses in Fig.~\ref{fig:refSpectrum}d). We use a nonlinear fitting routine to fit the center, width, amplitude, and noise-floor of a Gaussian distribution to the measurements. 

In Fig.~\ref{fig:vzDeflection} the intensity dependence of the center of the distribution, $dN/dp_z$, is shown for multi-photon ionization in 800~nm and 1400~nm of neon (circles) and argon (triangles, crosses). The intensity dependent momentum transfer has a different slope for the two wavelengths as is expected from the wavelength dependence of the asymptotic electron kinetic energy $K =  q^2 E_0^2/\left(2 m \omega^2\right)$ \cite{Corkum1989}.  Generally, with increasing intensity the center of $dN/dp_z$ is increasingly shifted in the direction of propagation. The upper limit of intensity arises because of saturation of single ionization, when a further increase of pulse energy does not result in an increase in the intensity seen by the neutral atoms. The actual saturation intensity depends on the ionization potential and the pulse duration (70~fs at 1400~nm, and 15~fs at 800~nm). 

In the following we will analyze the transfer of photon momentum to a free electron in the classical description of the light field \cite{Corkum1989}, using the fact that the photon momentum can be described by the Lorentz force \cite{Walser2000PRL}. A free electron is driven by the laser electric field (propagating along z)  $\mathbf{E} (z, t) = E_0 g \left(z,t\right) \left[ cos (\omega t - k z) \ \hat{e}_x + \sin (\omega t - k z) \ \hat{e}_y \right]$  with the pulse envelope $ g \left(z,t\right) = \exp \left( - \frac{ 2 \ln 2}{c^2 \tau^2} \left( z - c t \right)^2 \right)$, $\tau$ being the pulse length FWHM in intensity. With the magnetic field $\mathbf{B} = \hat{e}_z \times \mathbf{E}/c$ the equation of motion for the free electron is  $\mathbf{F} = q \mathbf{E} + q \mathbf{v} \times \mathbf{B}$. At non-relativistic velocities the magnetic field and the slowly varying pulse envelope can be treated as perturbations on the electron's motion. To first order the electron's velocity $\mathbf{v}$ only depends on the electric field. The second order correction for the momentum component $p_z$ includes the contribution of the magnetic field, $\mathbf{B}$, and the ponderomotive gradient $\mathbf{\nabla} U_p$. The action of the magnetic field generates a net forward momentum. The gradient of the pulse envelope results in a net negative force as the pulse passes over the electron.
\begin{equation}
 \frac{\ud p_z^e }{\ud t} = (q \mathbf{v} \times \mathbf{B}) \hat{e}_z - \frac{\partial}{\partial z} U_p  \quad.
 \label{eq:emotion2}
\end{equation}
Note, that $\mathbf{B}$ is for a constant pulse envelope, whereas the second term $-\partial_z U_p$ includes the pulse envelope.  For intensities in the tunneling regime the atom is most likely to ionize around the peak of the envelope (t=0). Since the pulse is short we can set $z = 0$ during the time the electron interacts with the laser. Using this simplification, Eq.~\ref{eq:emotion2} can be integrated to yield the net longitudinal momentum after the pulse has passed:
\begin{equation}
	p_z^e = \int_0^\infty \dot{p}_z^e (t) \ \ud t = \frac{q^2 E_0^2}{m \omega^2 c} \left( 1 - \frac{1}{2} \right) = \frac{K}{c} \quad .
	\label{eq:pz}
\end{equation}
The first term represents the forward momentum due to the radiation pressure on the free electron. The second term is the effect of the gradient in Eq.~\ref{eq:emotion2}. It removes the forward momentum associated with the ponderomotive energy. This is the reason $U_p$ does not appear in Eq.~\ref{eq:pz}. When the pulse has vanished there is still a forward momentum due to the electron's drift kinetic energy, $K$. We confirmed the validity of Eq.~\ref{eq:pz} using a numerical simulation to the complete Lorentz equation of motion including the pulse envelope. 

Eq.~\ref{eq:pz} agrees with the experimental data in Fig.\ \ref{fig:vzDeflection}. For comparison, in Fig.~\ref{fig:vzDeflection}, we also plot Eq.~\ref{eq:pz} plus a constant offset $I_p/c$. The dashed line, which includes this offset, overestimates the measured $p_z^e$ well beyond our experimental uncertainty. Thus, we conclude that the portion of the longitudinal momentum $I_p/c$, corresponding to the photons necessary to overcome the ionization energy, is not given to the photo-electron. It must be transferred to the centre of mass of the electron-ion system ($p_z^i=I_p/c$).

The agreement between Eq.~\ref{eq:pz} and the measurement allows us to draw another conclusion. It shows that the ponderomotive energy does not transfer any net momentum to the electron. This is because the ponderomotive term in Eq.~\ref{eq:emotion2} removes that part of the longitudinal momentum as the pulse passes the electron. The plasma-induced phase modulation of the pulse \cite{Yablonovitch1988} returns the momentum $p^L=U_p/c$ to the photon field. If the ponderomotive energy were necessary, the slope in Fig. \ref{fig:vzDeflection} would be different by a factor of two -- well outside the error bars. This conclusion means that linearly and circularly polarized pulses will transfer very different longitudinal momentum to the photoelectrons. In linearly polarized light most of the energy  absorbed by the electron upon ionization is ponderomotive energy and is therefore returned to the field. In circularly polarized light, half the absorbed energy is in $U_p$ and half in $K$. The momentum associated with $K$ remains with the photoelectron $p_z^e$. 

\begin{figure}
  \includegraphics[width=0.8\columnwidth]{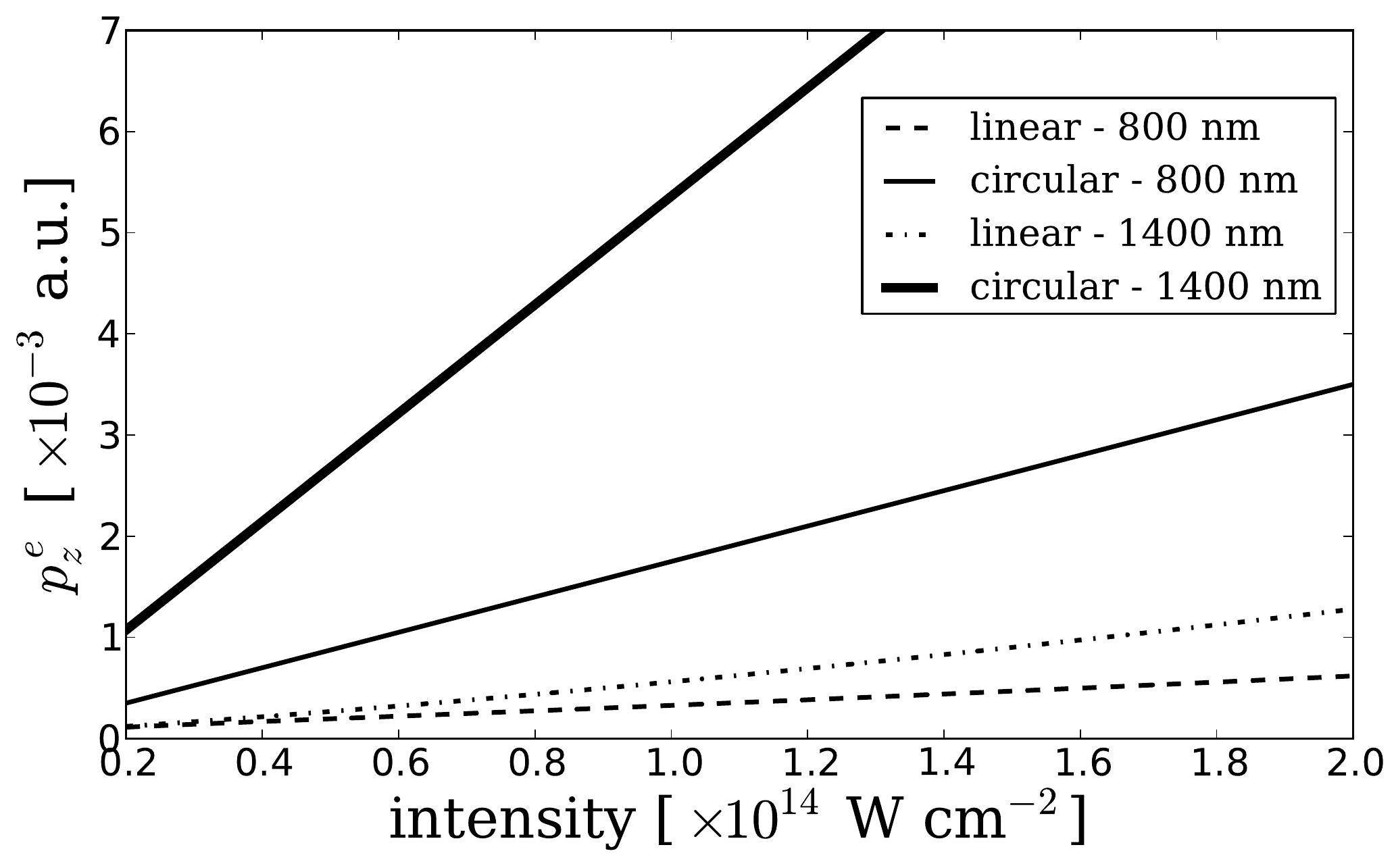}
  \caption{Net photoelectron longitudinal momentum ($p_z^e$) vs laser intensity calculated for linear and circularly polarized light. The curves are calculated from Eq.~\ref{eq:pz} using the kinetic energy averaged over one optical cycle.}
  \label{fig:calcMomentum}
\end{figure}

In Fig.~\ref{fig:calcMomentum} we compare the photoelectron longitudinal momentum calculated for circularly and linearly polarized light at two different wavelengths. In both cases we have used Eq.~\ref{eq:pz} to calculate the longitudinal momentum. Circular and linear polarization have very different kinetic energy distributions and this affects the net longitudinal momentum of the photoelectrons when averaged over one optical cycle. In circularly polarized light, the ionization probability is independent of the phase of the field. In linear polarization, the ionization probability and the kinetic energy are strongly depend on the phase. An electron ionized at the phase $\omega t_i$ has drift kinetic energy $K = q^2 E_0^2 \sin^2 \omega t_i / \left( 2 m \omega^2 \right)$. The average kinetic energy in linearly polarized light is $\bar{K} = \int_{-\pi/\omega}^{\pi/\omega} K \left( t_i \right) \frac{\ud N}{\ud t_i} \; \ud t_i$
where $t_i$ is the time of ionization and $\ud N/\ud t_i = W \left[ E\left(t_i\right) \right]$ is the Yudin-Ivanov sub-cycle ionization rate \cite{Yudin2001a}. In general $\bar{K}$ in linearly polarized light is much less than in circular polarization at a constant intensity. 

There are two important implications of our results. First, there has been a debate about the proper role of tunneling in multi-photon ionization \cite{Blaga2009, Reiss2008}. It is clear that tunneling -- a DC phenomenon -- provides no mechanism to transfer linear momentum to the ion. Although tunneling is a tremendous computational tool, our experiment shows that it remains an approximation. To our knowledge, how the momentum $I_p/c$ is shared between the electron and ion has never been addressed before in experiments, simulations, or models. 

Second, the photon momentum transfer to the electron has been invoked to account for THz generation in filaments \cite{Cheng2001,Amico2008,Zhou2011}. In the low fields typical of atmospheric gas breakdown, the momentum transfer will be very small for linearly polarized light. Our results suggest this can be augmented by using circular polarization and longer wavelength.

We are pleased to acknowledge the support of AFOSR, MURI grant W911NF-07-1-0475, CIPI and NSERC.


\end{document}